\documentclass[a4paper,11pt]{article}
\pdfoutput=1 

\usepackage{jinstpub}

\usepackage{dcolumn}
\usepackage[T1]{fontenc}
\usepackage[utf8]{inputenc}
\usepackage{tabularx}
\newcolumntype{q}{>{\hsize=1.6\hsize\linewidth=\hsize}X}
\newcolumntype{e}{>{\hsize=.4\hsize\linewidth=\hsize}X}
\usepackage{booktabs}
\usepackage[separate-uncertainty=true]{siunitx}
\usepackage{xcolor}
\usepackage{listings}
\usepackage{xifthen}
\usepackage{fancyvrb}

\definecolor{codeblue}{cmyk}{0.93,0.65,0.01,0}
\definecolor{codegray}{cmyk}{0.30,0.25,0.15,0.50}
\definecolor{codepurple}{cmyk}{0.10,0.80,0,0}
\definecolor{backcolour}{cmyk}{0.04,0.03,0.03,0}
\definecolor{codelightblue}{cmyk}{0.70,0.20,0,0}
\lstdefinestyle{mystyle}{
    backgroundcolor=\color{backcolour},
    commentstyle=\color{codelightblue},
    keywordstyle=\color{codeblue},
    numberstyle=\tiny\color{codegray},
    stringstyle=\color{codepurple},
    basicstyle=\ttfamily\footnotesize,
    breakatwhitespace=false,
		fancyvrb=true,
    breaklines=true,
    captionpos=b,
    keepspaces=true,
    numbers=left,
    numbersep=5pt,
    showspaces=false,
    showstringspaces=false,
    showtabs=false,
    tabsize=2
}
\newcommand\gtwo[2][]{\ifthenelse{\isempty{#2}}{$g^{(2)}(0)$}{\ifthenelse{\isempty{#1}}{$g^{(2)}(0)=#2$}{$g^{(2)}(0)=#2\pm#1$}}}
\newcommand\gtwot{$g^{(2)}(t)$}
\newcommand\name[1]{\textit{#1}}
\newcommand\ps[1]{\SI{#1}{\pico\second}}

\title{Efficient and Versatile Toolbox for Analysis of Time-Tagged Measurements}

\author[a,b,i]{Z. Lin,\note{Authors contributed equally.}}
\author[a,i,ii]{L. Schweickert,\note{Corresponding author.}}
\author[a]{S. Gyger,}
\author[a,iii]{K. D. J\"ons,\note{Current affiliation: Department of Physics, Paderborn University, 33098 Paderborn, Germany}}
\author[a]{and V. Zwiller}

\affiliation[a]{Department of Applied Physics, Royal Institute of Technology, Roslagstullsbacken 21, 114 21 Stockholm, Sweden}
\affiliation[b]{School of Precision Instrument and Opto-electronics Engineering, Tianjin University, Weijin Road 92, Tianjin, China}

\emailAdd{lucassc@kth.se}

\abstract{Acquisition and analysis of time-tagged events is a ubiquitous tool in scientific and industrial applications. With increasing time resolution, number of input channels, and acquired events, the amount of data can be overwhelming for standard processing techniques. We developed the \textbf{E}xtensible \textbf{T}ime-tag \textbf{A}nalyzer (ETA), a powerful and versatile, yet easy to use software to efficiently analyze and display time-tagged data. Our tool allows for flexible extraction of correlation from time-tagged data beyond start-stop measurements that were traditionally used. A combination of state diagrams and simple code snippets allows for analysis of arbitrary complexity while keeping computational efficiency high.}


\begin{document}
\maketitle
\flushbottom

\section{Introduction}
Extracting correlation from time resolved data~\cite{Priestley:1982} gives insights into the dynamics of a system under study over more than 18 orders of magnitude, from picoseconds to hours, in a single experiment. This makes it one of the most powerful tools for data analysis widely used in the sciences. Time resolution better than 8\,picoseconds is already available at the time of writing~\cite{EsmaeilZadeh.Los.ea:2020} and can be expected to reach the femtosecond range in the near future~\cite{Korzh.Zhao.ea:2020}. In optics, the correlation among photon detection events is often analyzed to investigate underlying physical processes~\cite{Bollinger.Thomas:1961, Kapusta.Wahl.ea:2015}. Examples include (i) light detection and ranging (LIDAR), where time-of-flight measurements, a subclass of correlation measurements, provide the distance to a reflective or scattering medium~\cite{Shan.Toth:2018}, (ii) random number generation where the timing and probability of the events generate random values~\cite{Herrero-Collantes.Garcia-Escartin:2017}, and (iii) characterization of quantum emitter properties and determination of the number of emitters under study~\cite{Chunnilall.Degiovanni.ea:2014}.
Correlation measurements are also required to characterize entangled states, be it well-known two-photon entangled states~\cite{Kocher.Commins:1967, Shih.Alley:1988}, more complex multi-photon entangled states such as GHZ~\cite{Greenberger.Horne.ea:1989} or cluster states~\cite{Briegel.Raussendorf:2001}.

\begin{figure}[htb]
    \centering
    \includegraphics[width=\textwidth]{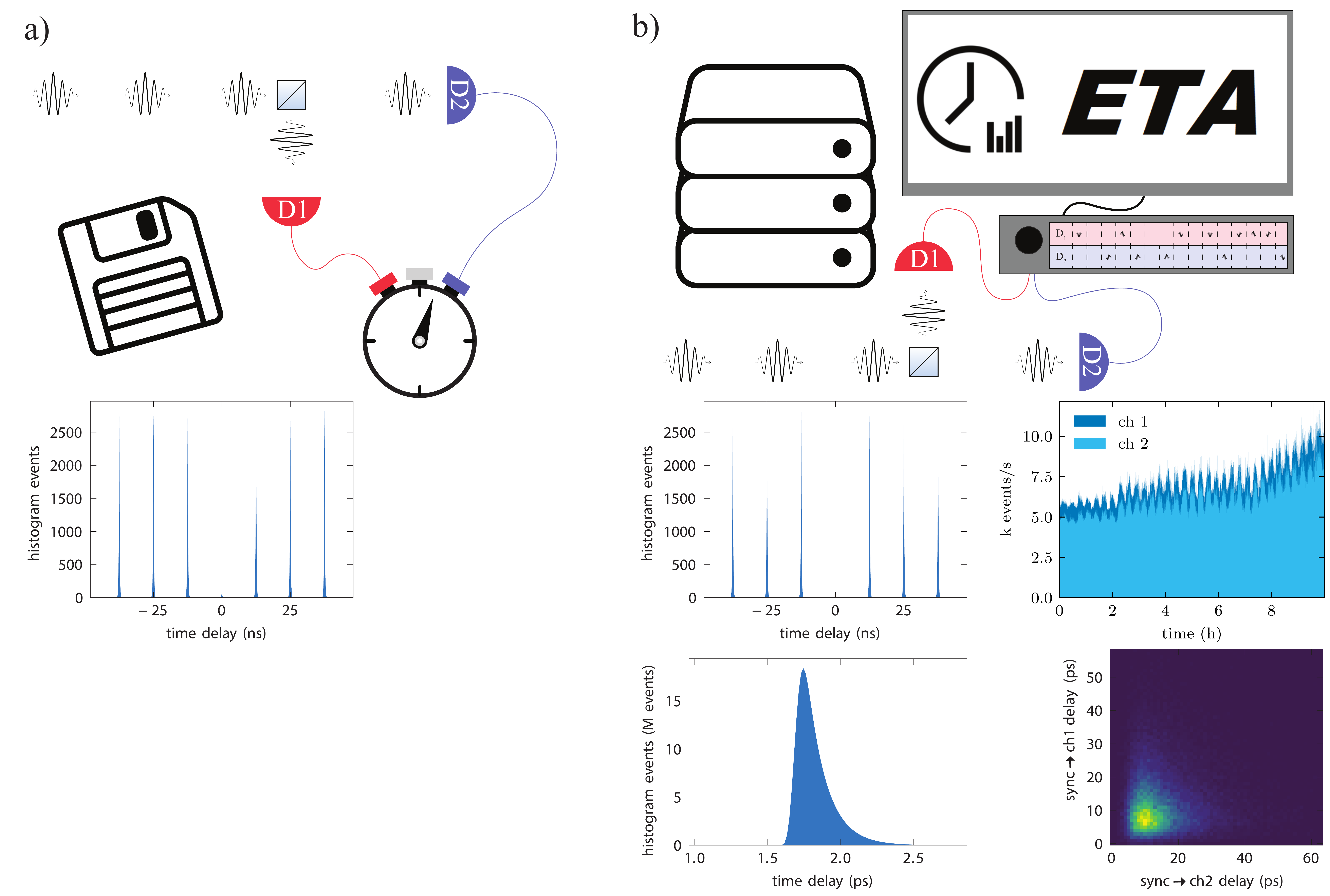}
    \caption{Configuration and results for time correlation measurements with (b) and without (a) time-tagging. (a) Correlating photons in time-to-analogue hardware yields the analyzed data directly with little storage space used. (b) When time-tagging photon arrival times, many different analysis methods can be applied to data from one experiment but a large amount of data is produced. Results depicted are: Correlation, count rate, lifetime, and two-time correlation.}
    \label{fig:old_vs_new}
\end{figure}

As an example, the usual experimental setup in quantum optics is based on the well-known Hanbury Brown and Twiss (HBT) experiment~\cite{Brown.Twiss:1956}, schematically shown in Figure~\ref{fig:old_vs_new}: a stream of photons is directed at a beam splitter with click detectors at each output. Here, the use of a beam splitter allows for detection events to be obtained at shorter time intervals than the detectors’ dead time. Using a single detector limits the time resolution of the system to the detector dead time but can still reveal a correlation on slower time scales~\cite{Steudle.Schietinger.ea:2012}.

Correlation between two click-detectors was historically measured with a time-to-amplitude converter, where one detector starts a timer and the other detector stops it again, generating a time interval value, as illustrated in Figure~\ref{fig:old_vs_new}a. After accumulating a significant number of them, these time intervals can be plotted in a histogram. The recent advent of time-tagging techniques~\cite{Roberts.Ali-Bakhshian:2010} for photon detection events with timing resolution comparable to the coherence and lifetimes of quantum emitters offers an alternative to the well established start-stop histograms obtained directly with analogue timing electronics. In time-tagging, fast electronics record the occurrence of each detection event with respect to an absolute time $t_0$, generating a timetable of events from all detectors, as illustrated in Figure~\ref{fig:old_vs_new}b. Rudimentary software offerings from manufacturers of these time-tagging devices, however, often require that the analysis method is selected from a predefined list of options ahead of time and only the resulting histogram is stored. This approach does not allow for more complex experiments, like e.g. entanglement swapping~\cite{Zopf.Keil.ea:2019, BassoBasset.Rota.ea:2019}, quantum key distribution~\cite{Ursin.Tiefenbacher.ea:2007}, continuous variable entanglement~\cite{Shalm.Hamel.ea:2013}, spin-spin entanglement~\cite{Delteil.Sun.ea:2016} or teleportation~\cite{Reindl.Huber.ea:2018}, and only a single method of analysis can be chosen per experiment.

When making full use of modern time-tagging hardware, instead of committing to an analysis method before the start of the experiment, all timing information can be saved to disk. The resulting time-tagged files can be analyzed after the measurement in various ways to extract correlation between the recorded channels. Depending on the experiment, time-tagged files can require terabytes of storage space. This makes data analysis a major hurdle and specialized software is needed to extract useful information in a reasonable time.
Therefore, efficient correlation extraction and processing of large data sets is still a widely encountered challenge in a broad range of applications and research fields.

The importance of user-defined analysis using time-tagged data becomes apparent when looking at an example in more detail:
When logging the arrival times of single photons emitted by a quantum dot along with the laser excitation events, we can extract the exciton lifetime, the biexciton lifetime, the exciton emission auto-correlation, the biexciton auto-correlation, the time evolution of count rates for exciton and biexciton, two-time correlation as well as cross-correlation between exciton and biexciton. Using time-tagging, these results can be extracted by analyzing data from only one experiment, not only saving time but also offering more reliable results since the data were acquired under identical conditions~\cite{Scholl.Hanschke.ea:2019}.

We created a versatile toolbox, \textbf{E}xtensible \textbf{T}ime-tag \textbf{A}nalyzer (ETA),\footnote{https://timetag.github.io/} for analysis of time-tagged data, enabling extraction of a wide range of information from one experiment with high efficiency.

So far, a modeling language designed to intuitively allow the specification of a particular time-tag analysis method to be executed by software has been elusive. Researchers tend to use a familiar general-purpose programming language to solve the problem at hand. The result is a clustered landscape of very specialized analysis scripts. With ETA, the user specifies the desired analysis method in a declarative style with a combination of graphical and traditional programming. Automatically selecting an appropriate algorithm, a just-in-time compiler combines these two inputs into an intermediate representation, which is then compiled into assembly code optimized for the target computer’s architecture. This procedure optimizes for fast analysis of large time-tag files at the cost of some upfront compilation time, while still maintaining flexibility.

While several software solutions for the analysis of time-tagged data have been made available~\cite{PicoQuantGmbH:2020}, ETA offers four key advantages:

\begin{enumerate}
    \item ETA uses an optimized compiler to automatically provide short analysis times. This allows for correlation histograms to be viewed in real-time during the measurement. Direct feedback for alignment or data preview is a sought-after feature commonly missing for time-tagging since the data acquisition rate is too large to handle for standard software.
    \item With our \name{Instrument Designer}, users can define the desired analysis method in a straight forward way by drawing state diagrams. Built-in functions and Python code can be executed whenever a new state is reached. This combination of state diagrams with custom code execution balances ease-of-use and flexibility when designing complex analysis methods.
    \item We disentangle the experiment from vendor specific code by providing a unified user interface for the analysis of time-tagged data from all major time-to-digital converters.
    \item ETA is open source and designed to grow with the help of the scientific community to rise to the challenges of the future.
\end{enumerate}

We expect ETA to find uses in studies of single quantum emitters like atoms and molecules~\cite{Aharonovich.Englund.ea:2016, Becker:2012}, LIDAR~\cite{Buller.Wallace:2007}, quantum entanglement~\cite{Freedman.Clauser:1972,DeGreve.Yu.ea:2012,Gao.Fallahi.ea:2012} and fluorescence correlation spectroscopy measurements~\cite{Bohmer.Pampaloni.ea:2001}, as well as quantum key distribution protocols~\cite{Bennett.Brassard:1984,Yin.Li.ea:2020} where data from remote detectors needs to be synchronized and correlated. 


\section{Software description}
\begin{figure}[htb]
    \centering
    \includegraphics[width=\textwidth]{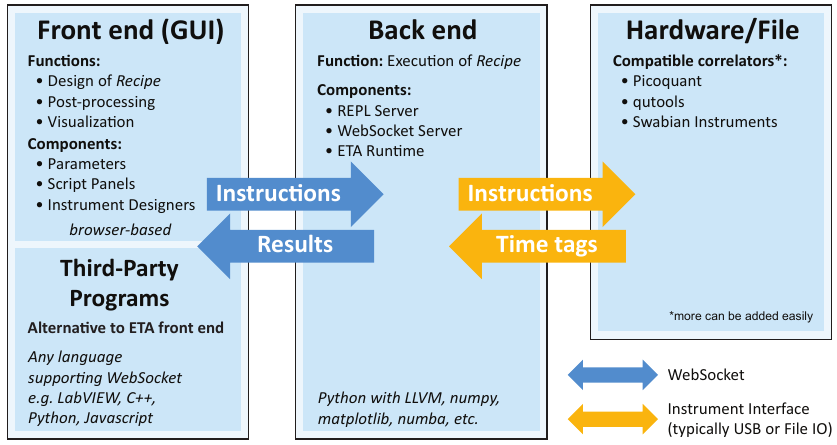}
    \caption{Software architecture of ETA. Time recording devices interact with the ETA back end via an interface like USB or via a saved file. The ETA back end receives its instructions from the ETA front end. Since the WebSocket protocol is used for the communication channel between back end and front end, a user-developed program can easily replace or extend the included front end, allowing for integration into an existing ecosystem.}
    \label{fig:architecture}
\end{figure}

ETA combines a high-performance back end and a flexible, intuitive graphical user interface. The front end is used to design an analysis method in a graphical programming workflow. As illustrated in Figure~\ref{fig:architecture}, it sends instructions to the back end, which compiles these instructions into optimized code and performs the analysis of the time-tagged data. The result is sent back to the front end for post-processing and displaying of the result. The user can describe how data is analyzed by creating a \name{Virtual Instrument}: After entering the \name{Instrument Designer} environment, a state diagram can be drawn where events, read from a time-tag file or even created on-the-fly, cause transitions among states. Upon arrival to a state or invocation of a specific transition, a user-defined action can be triggered. The specifics of the action are described in the coding panel on the right-hand side of the \name{Instrument Designer} using Python-like syntax. Multiple \name{Virtual Instruments} can be combined and used from within a \name{Script Panel}. There, additional data processing, analysis and plotting can be performed on the histogram calculated by the ETA back end. The programming language chosen for the \name{Script Panel} is Python.
Standard functionality is provided for, amongst others, lifetime, correlation and count rate --- all possible in real-time (see also section~\ref{sec:examples}).
In case a more specialized analysis method is desired, the user can build custom functionality by using provided functions or embedded code blocks for fully customized analysis.

\subsection{Software Architecture}

Due to its division into front end and back end, ETA can be used in a multitude of ways, resulting in both cross-platform and cross-device compatibility. A WebSocket-based protocol is used for communication between front end and back end, allowing the computational power of the back end to be easily integrated into existing software. The front end is based on web-technologies to offer familiar aesthetics, displayed in a standalone software. A web-hosted version is available for mobile devices that support a browser. The back end installs as a Python package and provides a library interface, allowing integration with an individual Python workflow and easy installation across Microsoft Windows, Mac OS X, and Linux.

\subsection{User Interface (Front end)}
\begin{figure}[htb]
    \centering
    \includegraphics[width=\textwidth]{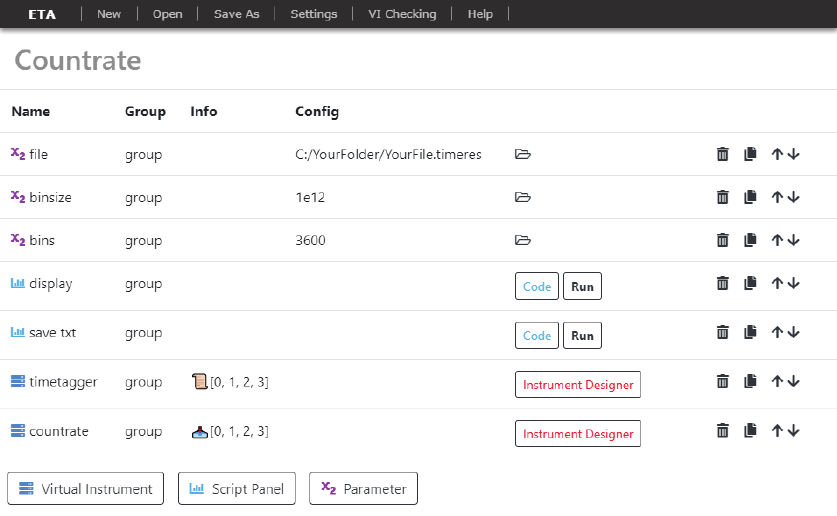}
    \caption{The \name{Main Panel} of ETA's front end with a simplified count rate \name{Recipe} loaded. The list of components shows three \name{Parameters}, two \name{Script Panels}, and two \name{Virtual Instruments}. A \name{Script Panel} can be opened by clicking the associated \name{Code} button. An \name{Instrument Designer} can be opened by clicking the corresponding button of the \name{Virtual Instrument}.}
    \label{fig:GUIexample}
\end{figure}
\subsubsection*{Main Panel}
The main panel, shown in Figure~\ref{fig:GUIexample}, consists of a list composed of elements of one of three types:
\begin{itemize}
    \item \textbf{\name{Parameters}} are strings that can be defined by the user on the \name{Main Panel} and interpreted as Boolean, integer, float or string in the \name{Instrument Designer} or \name{Script Panel}. In most included \name{Recipes} a \name{Parameter} is used to define the path to the time-tagged data.
    \item \textbf{\name{Virtual Instruments}} are at the heart of ETA. With the \name{Instrument Designer} instructions can be laid out inside the \name{Virtual Instruments} on how ETA analyzes time-tags. Each source of time-tags, e.g. a measurement device, is represented as a \name{Virtual Instrument}. This allows the combination of time-tags from multiple devices. A delay line or a virtual beam splitter (see Section~\ref{sec:examples}) are further examples of \name{Virtual Instruments}.
    \item \textbf{\name{Script Panels}} are text editors with syntax highlighting for the Python language and are used to manage the analysis. Here the user decides which files to read and which \name{Virtual Instruments} to use. Also, it offers the possibility to do post-processing, secondary analysis, and visualization of the results returned by the \name{Virtual Instruments}.
\end{itemize}

\subsubsection*{The Instrument Designer}
\begin{figure}[htb]
    \centering
    \includegraphics[width=\textwidth]{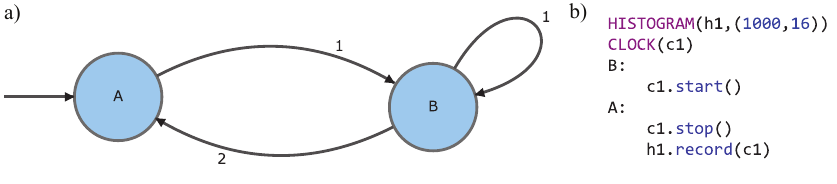}
    \caption{Lifetime analysis. (a) State diagram for lifetime analysis.  (b) \name{Actions} and \name{Tools} for lifetime analysis.}
    \label{fig:lifetime_states}
\end{figure}

The \name{Instrument Designer} is divided into a state diagram on the left-hand side, e.g. Fig.~\ref{fig:lifetime_states}a, and the \name{Actions} and \name{Tools} panel on the right-hand side , e.g. Fig.~\ref{fig:lifetime_states}b. In the state diagram, blue circles represent states. They can be placed by double-clicking or by dragging out from an existing circle and can be named by double-clicking into the circle. 
There must be a single state where the analysis starts, which can be defined by selecting a state and pressing Shift+I (Initial). Connections between states, i.e. transitions (arrows), can be created by dragging out from an existing state onto another or onto itself. If a state is created by dragging out from an existing circle onto nothing, they will also be linked by a transition. Each transition must be labeled with all channel numbers that trigger this transition separated by commas. This can be done by double-clicking the transition.
The \name{Actions and Tools} panel on the right-hand side defines what happens when a certain transition is triggered. This if-statement is represented by a description of the trigger followed by a colon and an indented block describing the Action. The trigger description on the \name{Actions} and \name{Tools} side can be automatically created by clicking on a transition or a state and pressing Shift+T (Trigger) and works as detailed in table~\ref{tab:triggers}.
\begin{table}[htb]
    \centering
    \caption{Trigger declaration. Triggers are declared in the script panel of the \name{Instrument Designer}.}
    \begin{tabularx}{\textwidth}{eq}
        trigger declaration  & meaning \\
        \midrule
        A:                      & When state A is reached. \\
        -{}-2-{}->A:            & When state A is reached via an event on channel 2.\\
        B-{}-1,2-{}->A:         & When state A is reached from state B via an event on channel 1 or channel 2.\\
        \bottomrule
    \end{tabularx}
    \label{tab:triggers}
\end{table}

To perform \name{Actions} when a Trigger fires, often a \name{Tool} must be defined with which the \name{Action} can be performed. The most commonly used Tools are clocks and histograms which can be created by writing \lstinline{CLOCK(clock_name)} and \lstinline{HISTOGRAM(histogram_name, number_of_bins, width_of_bins)}, respectively. A clock can then be started and stopped with \lstinline{clock_name.start()} and \lstinline{clock_name.stop()}, while the recorded time difference can be entered into the histogram with \lstinline{histogram_name.record(clock_name)} after the clock has been stopped. These \name{Actions} must be placed in the indented block of a trigger. Notably, the width of the bins can be set individually, thereby allowing e.g. logarithmic bins for capturing fast and slow processes in the same evaluation.

\subsection{Execution of Analysis Recipe (Back end)}
ETA is fast, efficient and compatible with a growing number of time-to-digital converters including PicoQuant, qutools, and Swabian Instruments. The compatibility with different file formats is provided through a flexible data loading layer. Below we highlight some of the algorithms and design choices responsible for ETA's efficiency and speed. 

\subsubsection*{Algorithms} 
ETA automatically generates a highly optimized data processing program based on the analysis instructions defined in the recipe using both state diagram and \name{Tools} and \name{Actions} in the \name{Instrument Designer}. ETA accomplishes this by selecting and combining algorithms appropriately. 

An N-way tournament sort algorithm~\cite{Knuth:1968} is used in the virtual channels, leveraging the fact that the time-tags are already pre-sorted in every channel, and most of the analysis methods are order-preserving. This means an event going into the delay line first comes out first, even if the absolute timing was changed inside the delay line. Based on this observation, using an N-way tournament sort can achieve a speedup of the time-tag analysis from a computational complexity $O(m \log(m))$ to $O(m \log(n))$, where $m$ is the total number of events and $n$ is the number of detector channels, compared to the Quicksort algorithm~\cite{Hoare:1961}, a fast algorithm in the general case. This results in a speedup of  $\frac{\log(m)}{\log(n)}$, which is typically 17x when doing correlation on a 1 GB time-tag file. 

When performing correlation analysis, a ring buffer algorithm~\cite{Bischof:2012, Ballesteros.Proux.ea:2019} is used to reduce computational complexity from $O(m^2)$ to $O(m k)$, where $m$ is the total number of events and $k$ is the average number of two-channel correlation counts within the maximum time delay in the histogram. This results in a speedup of $\frac{m}{k}$, which can be several orders of magnitudes for large time-tag files with high event rates.

If the default choice of algorithm for implementing the \name{Tools} and \name{Actions} does not suit the user's needs, an embedded block of code, which can be written in Python, can also be used in the \name{Tools} and \name{Actions} panel. All of this code will then be combined and converted by Numba~\cite{Lam.Pitrou.ea:2015}, a static Python compiler, into LLVM~\cite{Lattner.Adve:2004} code and afterwards into fast machine assembly.

\subsubsection*{Optimization} 

Unlike most of the existing data analysis tools, which become unwieldy and progressively slower as features are added, ETA uses a just-in-time compilation method. This allows ETA to compile only the algorithmic methods required for the current analysis, resulting in optimized native machine assembly code. Internally, ETA utilizes LLVM, a state-of-the-art assembly code generator and optimizer, to generate an intermediate representation for the analysis defined in the recipe. This takes into account the input time-tag file format and the variables defined in the \name{Virtual Instruments}. Variables are converted to constants before execution whenever possible to reduce the number of instructions at run-time. The intermediate representation is then translated using optimization tricks such as branch table generation and function-call elimination. This yields fast assembly code that runs directly on the target-CPU without an interpreter or virtual machine, resulting in performance similar to optimized C/C++ code for a specific type of analysis, while still maintaining flexibility via the \name{Instrument Designer}. Adding features does therefore not require a re-write of the program and a hard-to-manage code base with several similar analysis-specific functions that would have to be selected via computationally costly if-statements.

\section{Illustrative examples}
\label{sec:examples}
\subsection{Lifetime, Start-stop, and Correlation analysis}

Performing a lifetime analysis means sorting time differences between a synchronization event (sync) and the detection event, e.g. the arrival time of a photon. For a lifetime analysis, the state diagram has two states, e.g. A and B, where a transition from A to B starts a clock to measure the time difference and a transition from B to A stops this clock. The $\text{A}\rightarrow\text{B}$ transition (start) therefore occurs with an event on the sync channel and the $\text{B}\rightarrow\text{A}$ transition (stop) occurs with an event on the channel under investigation. If some photons are lost between source and detector, which is typically the case, several sync events can occur consecutively. To record only the shortest time differences, it is necessary to restart the clock on each sync event. We therefore, draw another transition from state B to itself, labeled with the sync channel number. We then trigger the Action \lstinline{c1.start()} with the trigger \lstinline{B:}, i.e. whenever we enter state B either from state A or from itself. And we trigger the \name{Actions} \lstinline{c1.stop()} and \lstinline{h1.record(c1)} with the trigger \lstinline{A:}, i.e. whenever we enter state A, in this case only from state B.
These are all the instructions specifying how ETA's back end analyzes the time-tagged data.

To load the measurement data we need to create a representation of the hardware on the \name{Main Panel} of the front end. Therefore, we create another \name{Virtual Instrument}, enter the \name{Instrument Designer} and specify the name of the source and number of channels that should be read from the file with \lstinline{RFILE(timetagger_name,[0,1,2])}. 

We then create a \name{Script Panel} which already includes the minimum example required to save the histogram to file:
\begin{lstlisting}[language=Python]
import numpy as np
cut=eta.clips("C:\\Path_to_file\\File.timeres")
result= eta.run({"timetagger_name":cut})
histogram = result["h1"] # get list from result
np.savetxt("h1.txt",histogram) # save txt file
\end{lstlisting}
The whole analysis can then be executed by clicking the "Run"-button of the \name{Script Panel}. A more sophisticated version of a \name{Recipe} for lifetime analysis is included with the software.

\label{sec:corr_anal}
\begin{figure}[htb]
    \centering
    \includegraphics[width=\textwidth]{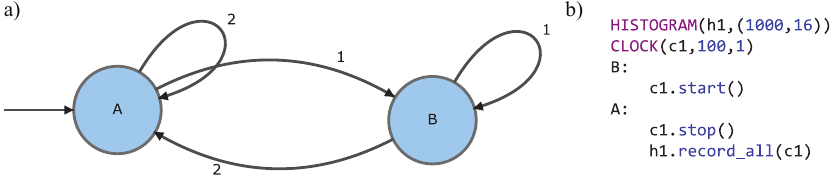}
    \caption{Correlation analysis. (a) State diagram for correlation analysis.  (b) \name{Actions} and \name{Tools} for correlation analysis.}
    \label{fig:correlation_states}
\end{figure}

Another prominent measurement, the second-order intensity time correlation function defined as $g^{(2)}(t)=\frac{\left<I(t)I(t+\tau)\right>}{\left<I(t)^2\right>}$ is essential in characterizing quantum sources: as an example, single photon emitters are identified with \gtwo{} values reaching well below 0.5~\cite{Kimble.Dagenais.ea:1977}. Auto-correlation measurements are also used in fluorescence correlation spectroscopy~\cite{Magde.Elson.ea:1972} and are often performed to identify the number of single-photon emitters, as well as diffusion and blinking times~\cite{Moerner.Fromm:2003, Dickson.Cubitt.ea:1997}.

Formerly, simple start-stop measurements, that can be obtained with analog time to amplitude converters, were used to extract a reasonable approximation~\cite{Davidson.Mandel:1968} that can yield the same result at $t=0$, the value that is often of interest.
In ETA, the start-stop \name{Recipe} can be easily made to simulate the behavior of the dedicated electronics in the old days, and it works similarly to the lifetime \name{Recipe}, by simply removing a reset transition.

\begin{figure}[htb]
    \centering
    \includegraphics[width=\textwidth]{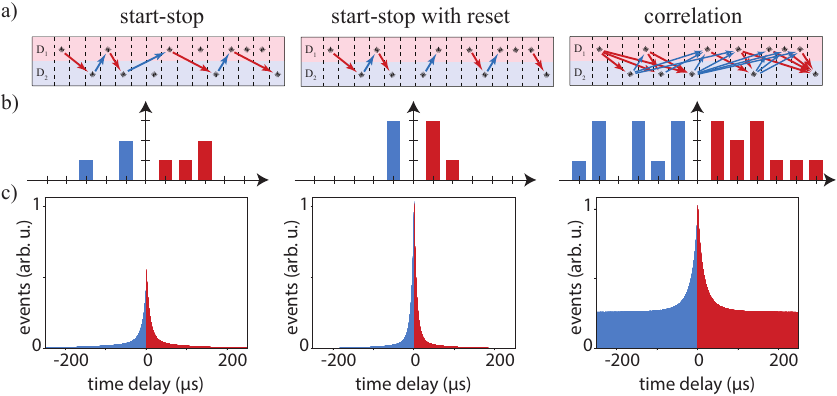}
    \caption{Comparison of different histogram calculations. Event streams (a), corresponding histograms (b), and exemplary data using fluorescence from a quantum dot (c) for, from left to right, start-stop analysis, start-stop with reset, and correlation with up to 6 neighbours.}
    \label{fig:ss_vs_corr}
\end{figure}

With ETA, a real \gtwot{} measurement which contains more information than a start-stop measurement (see Fig.~\ref{fig:ss_vs_corr}), can also be easily achieved with a correlation \name{Recipe}. 

Since ETA can automatically produce an optimized code that is fast enough to perform full correlation, the only change required for the user is to specify in the \name{Instrument Designer}, that time differences between all events, not only the neighbouring ones, have to be recorded into the histogram. This is done by allowing the clock to be started many times. Each started instance can be individually stopped when an event on the second channel is encountered. Since \name{Actions} have to take place when consecutive events occur on the same channel, both states A and B need a loop back to themselves. If we label them with the same channel numbers as the transitions pointing at them already, a second photon on the same channel will trigger another start of the clock in case of state B and a stop of the clock, as well as a recording to the histogram in case of state A. This will result in a correlation of the events on channel A with the events on channel B. A more sophisticated version of a \name{Recipe} for correlation analysis is included with the software.

Figure~\ref{fig:ss_vs_corr}a shows an example of a raw event stream. Arrows indicate the recorded time intervals for positive (red) and negative (blue) time delay. As described before, a start-stop measurement will consider each event only once. When resetting the start, the last event in a row of consecutive events on the start channel will be used while the previous ones will be discarded. Since we can access the full event stream when time-tagging, to perform correlation, we can reuse any detection event to record time differences with all other events into a histogram. This is not possible with a simple start-stop type measurement. In Figure~\ref{fig:ss_vs_corr}b we illustrate this with a correlation where the stopping event is at most 6 time slots away from the starting event. Figure~\ref{fig:ss_vs_corr}c shows a real-world example of data obtained from recording the arrival times of photons from a single quantum dot. The data in Figure~\ref{fig:ss_vs_corr}c has been normalized to the highest value in the correlation case for all three panels for easier comparison. A value of 1 does not represent the Poisson level in this case. While the result close to time delay 0 is similar between the start-stop evaluation with reset and the correlation evaluation, only correlation of all photons with each other provides accurate results for long time delays.

\subsubsection*{Real-time and multi-threading support}
ETA allows processing new data while displaying and updating the result of the already evaluated data, which we call real-time analysis.
Time-tag data can be streamed into ETA in segments for real-time analysis. Due to the nature of ETA, those features are implemented in the compiling stage and in a recipe-agnostic manner.
This means real-time analysis can be enabled simply via the \name{Script Panel} in any type of ETA \name{Recipe}. It is provided with the built-in lifetime and correlation \name{Recipes}. The data can either be read from a file, while it is still being written or can be read-in directly from the time-to-digital converter. ETA will perform the analysis by automatically pausing and resuming as new data become available. In both cases, either the full time-tag file or only the analyzed data can be stored. 
This mode is often used for direct evaluation of a single-photon detector in an oscilloscope-like fashion. With this, a single-photon detector can, in many scenarios, replace a high-speed photo-diode, an important advantage when measuring very weak light intensities.

Previously stored data files can also be cut into smaller segments for faster parallel processing in a MapReduce~\cite{Dean.Ghemawat:2008} style method, in which the segments are analyzed individually into histograms (map stage), and aggregated with a user-specified method, usually a simple sum or concatenate (reduce stage). This can be useful to generate a quick preview of the analysis result at the cost of losing correlations between events across different segments.

\subsection{Simulation}
\label{sec:simulation}
\begin{figure}[htb]
    \centering
    \includegraphics[width=\textwidth]{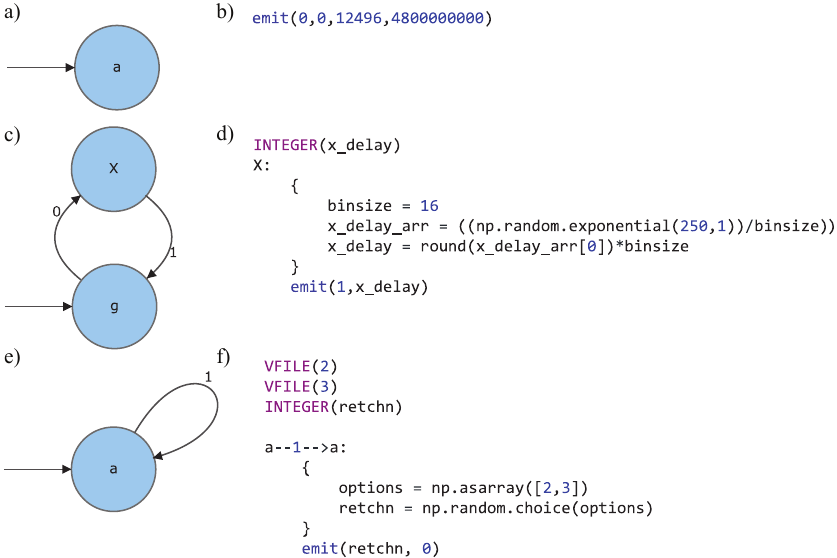}
    \caption{Simulation of time-correlation data. (a) State diagram for the generation of the sync with just the minimum requirements. (b) \name{Actions} and \name{Tools} panel for the generation of the sync with \lstinline{emit(channel,delay,repetition_time,number_of_repetitions)}. (c) State diagram for the simulation of a single-photon emitter. (d) \name{Actions} and \name{Tools} for a delay randomly picked from an exponentially decaying distribution. (e) State diagram for a beam splitter triggered by the decay of the single-photon emitter. (f) \name{Actions} and \name{Tools} for the simulated beam splitter with a 50:50 choice of which output channel is used.}
    \label{fig:simulation_states}
\end{figure}
By using the \lstinline{emit()} function in the \name{Instrument Designer} it is possible to create a custom time-tag file in memory or even on disk. To showcase this functionality, we create a stream of separated events suitable to simulate a pulsed \gtwot{} measurement with \SI{100}{\percent} efficiency. The first \name{Virtual Instrument}, shown in Figure~\ref{fig:simulation_states}a and \ref{fig:simulation_states}\,b, contains just one state with an initial-state marker, since that is the minimum requirement, and uses \lstinline{emit(0, 0, 12496, 4.8E9)} to create an event on channel\,1 with \ps{0} delay every \ps{12496}, $4.8\times10^{9}$ times, resulting in \SI{60}{\second} of \SI{80}{\mega\hertz} sync pulses. We then create a second \name{Virtual Instrument}, shown in Figure~\ref{fig:simulation_states}c and \ref{fig:simulation_states}d, with two states: a state called "g", representing the ground state and a state called "X" representing the excited state. On an event on channel\,0, the ground state can be excited and upon arriving at "X" a delayed emission on channel\,1 is triggered that will cause a return to the state "g". To generate this random delay, we make use of the embedded code block, where we can use all functions supported by the Numba compiler. By sampling from an exponential probability distribution, we can simulate the emission of a two-level system. \lstinline{emit(1,x_delay)} is then called to create a virtual stream of single-photon-like events. In a third \name{Virtual Instrument} we pick up this stream with a transition from a single state looping to itself, by choosing the virtual channel\,1 we just created. We then use another embedded code block to randomly choose between emitting on channel\,2 or 3 with equal probability to mimic a 50:50 beam splitter. We can now correlate channel\,2 and 3 as we did in section~\ref{sec:corr_anal}. A more sophisticated version of a quantum emitter simulation \name{Recipe} is included with the software.
\begin{figure}[htb]
    \centering
    \includegraphics[width=\textwidth]{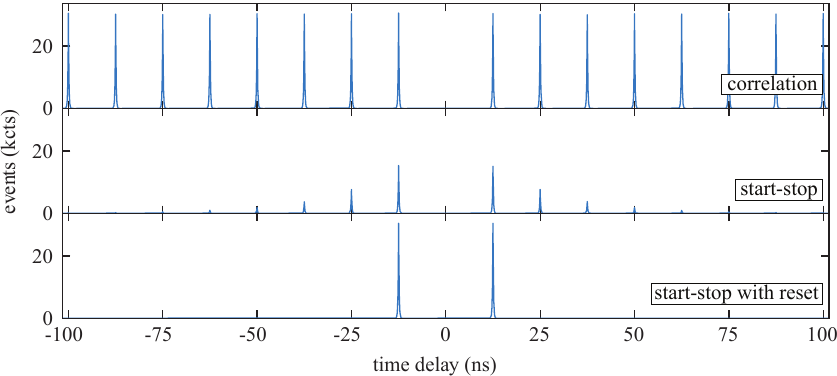}
    \caption{Simulated single-photon emission with \SI{100}{\percent} efficiency in generation, collection, routing and detection. The top panel shows a full correlation, the middle panel shows a start-stop analysis and the bottom panel shows a start-stop analysis where the start is reset when consecutive photon events are registered on the same channel.}
    \label{fig:simulation_result}
\end{figure}
Figure~\ref{fig:simulation_result} shows the result of this simulation, comparing correlation (top panel), start-stop analysis (middle panel) and start-stop analysis with reset of the clock upon consecutive events on the start channel (bottom panel). The start-stop method with reset only shows a single side-peak in each direction of the time delay since no event can be reused which would be necessary for a correlation analysis.

\section{Conclusion}
This software was developed with time-correlated single photon counting in mind. This technique is used in, among others, fluorescence microscopy and quantum optics, but is certainly not limited to these use cases. Analysis of time-tagged files instead of start-stop measurements allows for the extraction of as much information as possible from a single experiment, resulting in important time savings. Due to ETA's user-friendliness, it can also reduce the time spent on programming the analysis of recorded data. The program can perform novel analysis that are yet to be defined while remaining fast and robust. Large numbers of detectors, such as in Boson sampling~\cite{Wang.Qin.ea:2019} and ambitious linear optics quantum computation schemes~\cite{Knill.Laflamme.ea:2001} with associated large file sizes pose no problem for our software. The support of data from multiple time-taggers and from a multitude of vendors allows for use cases where the correlation of signals from remote sources, like in quantum key distribution, needs to be performed. Even simulation of time series data is possible due to the flexibility. A vast number of research fields could benefit from using our software due to time savings and unlocked potential.

\begin{acknowledgments}
This project has received funding from the European Union's Horizon 2020 research and innovation program under grant agreement No. 820423 (S2QUIP) and No. 899814 (Qurope), the Knut and Alice Wallenberg Foundation grant ”Quantum  Sensors”, the European Research Council (307687 (NaQuOp)), the Joint China-Sweden Mobility programme (STINT), the Swedish Research Council (VR) through the VR grant for international recruitment of leading researchers (Ref: 2013-7152), and Q-LID  Quantum Light Detectors (Ref: 2018-04251).
K.D.J. acknowledges funding from the Swedish Research Council (VR) via the starting grant HyQRep (Ref.: 2018-04812) and The Göran Gustafsson Foundation (SweTeQ).
\end{acknowledgments}

\section*{Data Sharing Policy}
Data sharing is not applicable to this article as no new data were created or analyzed in this study.

\bibliographystyle{unsrtnat}
\bibliography{references}

\begin{thebibliography}{46}
\providecommand{\natexlab}[1]{#1}
\providecommand{\url}[1]{\texttt{#1}}
\expandafter\ifx\csname urlstyle\endcsname\relax
  \providecommand{\doi}[1]{doi: #1}\else
  \providecommand{\doi}{doi: \begingroup \urlstyle{rm}\Url}\fi

\bibitem[Priestley(1982)]{Priestley:1982}
M.~B. Priestley.
\newblock \emph{Spectral Analysis and Time Series}.
\newblock Probability and Mathematical Statistics. {Elsevier}, {London},
  October 1982.
\newblock ISBN 978-0-12-564922-3.

\bibitem[Esmaeil~Zadeh et~al.(2020)Esmaeil~Zadeh, Los, Gourgues, Chang,
  Elshaari, Zichi, {van Staaden}, Swens, Kalhor, Guardiani, Meng, Zou,
  Dobrovolskiy, Fognini, Schaart, Dalacu, Poole, Reimer, Hu, Pereira, Zwiller,
  and Dorenbos]{EsmaeilZadeh.Los.ea:2020}
Iman Esmaeil~Zadeh, Johannes W.~N. Los, Ronan B.~M. Gourgues, Jin Chang, Ali~W.
  Elshaari, Julien~Romain Zichi, Yuri~J. {van Staaden}, Jeroen P.~E. Swens,
  Nima Kalhor, Antonio Guardiani, Yun Meng, Kai Zou, Sergiy Dobrovolskiy,
  Andreas~W. Fognini, Dennis~R. Schaart, Dan Dalacu, Philip~J. Poole,
  Michael~E. Reimer, Xiaolong Hu, Silvania~F. Pereira, Val Zwiller, and
  Sander~N. Dorenbos.
\newblock Efficient {{Single}}-{{Photon Detection}} with 7.7 ps {{Time
  Resolution}} for {{Photon}}-{{Correlation Measurements}}.
\newblock \emph{ACS Photonics}, 7\penalty0 (7):\penalty0 1780--1787, July 2020.
\newblock \doi{10.1021/acsphotonics.0c00433}.

\bibitem[Korzh et~al.(2020)Korzh, Zhao, Allmaras, Frasca, Autry, Bersin, Beyer,
  Briggs, Bumble, Colangelo, Crouch, Dane, Gerrits, Lita, Marsili, Moody,
  Pe{\~n}a, Ramirez, Rezac, Sinclair, Stevens, Velasco, Verma, Wollman, Xie,
  Zhu, Hale, Spiropulu, Silverman, Mirin, Nam, Kozorezov, Shaw, and
  Berggren]{Korzh.Zhao.ea:2020}
Boris Korzh, Qing-Yuan Zhao, Jason~P. Allmaras, Simone Frasca, Travis~M. Autry,
  Eric~A. Bersin, Andrew~D. Beyer, Ryan~M. Briggs, Bruce Bumble, Marco
  Colangelo, Garrison~M. Crouch, Andrew~E. Dane, Thomas Gerrits, Adriana~E.
  Lita, Francesco Marsili, Galan Moody, Cristi{\'a}n Pe{\~n}a, Edward Ramirez,
  Jake~D. Rezac, Neil Sinclair, Martin~J. Stevens, Angel~E. Velasco, Varun~B.
  Verma, Emma~E. Wollman, Si~Xie, Di~Zhu, Paul~D. Hale, Maria Spiropulu,
  Kevin~L. Silverman, Richard~P. Mirin, Sae~Woo Nam, Alexander~G. Kozorezov,
  Matthew~D. Shaw, and Karl~K. Berggren.
\newblock Demonstration of sub-3 ps temporal resolution with a superconducting
  nanowire single-photon detector.
\newblock \emph{Nature Photonics}, 14\penalty0 (4):\penalty0 250--255, April
  2020.
\newblock \doi{10.1038/s41566-020-0589-x}.

\bibitem[Bollinger and Thomas(1961)]{Bollinger.Thomas:1961}
L.~M. Bollinger and G.~E. Thomas.
\newblock Measurement of the {{Time Dependence}} of {{Scintillation Intensity}}
  by a {{Delayed}}-{{Coincidence Method}}.
\newblock \emph{Review of Scientific Instruments}, 32\penalty0 (9):\penalty0
  1044--1050, September 1961.
\newblock \doi{10.1063/1.1717610}.

\bibitem[Kapusta et~al.(2015)Kapusta, Wahl, and Erdmann]{Kapusta.Wahl.ea:2015}
Peter Kapusta, Michael Wahl, and Rainer Erdmann, editors.
\newblock \emph{Advanced {{Photon Counting}}: {{Applications}}, {{Methods}},
  {{Instrumentation}}}, volume~15 of \emph{Springer {{Series}} on
  {{Fluorescence}}}.
\newblock {Springer International Publishing}, {Cham}, 2015.
\newblock ISBN 978-3-319-15635-4 978-3-319-15636-1.
\newblock \doi{10.1007/978-3-319-15636-1}.

\bibitem[Shan and Toth(2018)]{Shan.Toth:2018}
Jie Shan and Charles~K Toth.
\newblock \emph{Topographic Laser Ranging and Scanning: Principles and
  Processing}.
\newblock {CRC Press, Taylor \& Francis Group}, 2018.
\newblock ISBN 978-1-315-15438-1.

\bibitem[{Herrero-Collantes} and
  {Garcia-Escartin}(2017)]{Herrero-Collantes.Garcia-Escartin:2017}
Miguel {Herrero-Collantes} and Juan~Carlos {Garcia-Escartin}.
\newblock Quantum random number generators.
\newblock \emph{Reviews of Modern Physics}, 89\penalty0 (1):\penalty0 015004,
  February 2017.
\newblock \doi{10.1103/RevModPhys.89.015004}.

\bibitem[Chunnilall et~al.(2014)Chunnilall, Degiovanni, K{\"u}ck, M{\"u}ller,
  and Sinclair]{Chunnilall.Degiovanni.ea:2014}
Christopher~J. Chunnilall, Ivo~Pietro Degiovanni, Stefan K{\"u}ck, Ingmar
  M{\"u}ller, and Alastair~G. Sinclair.
\newblock Metrology of single-photon sources and detectors: A review.
\newblock \emph{Optical Engineering}, 53\penalty0 (8):\penalty0 081910, July
  2014.
\newblock \doi{10.1117/1.OE.53.8.081910}.

\bibitem[Kocher and Commins(1967)]{Kocher.Commins:1967}
Carl~A. Kocher and Eugene~D. Commins.
\newblock Polarization {{Correlation}} of {{Photons Emitted}} in an {{Atomic
  Cascade}}.
\newblock \emph{Physical Review Letters}, 18\penalty0 (15):\penalty0 575--577,
  April 1967.
\newblock \doi{10.1103/PhysRevLett.18.575}.

\bibitem[Shih and Alley(1988)]{Shih.Alley:1988}
Y.~H. Shih and C.~O. Alley.
\newblock New {{Type}} of {{Einstein}}-{{Podolsky}}-{{Rosen}}-{{Bohm Experiment
  Using Pairs}} of {{Light Quanta Produced}} by {{Optical Parametric Down
  Conversion}}.
\newblock \emph{Physical Review Letters}, 61\penalty0 (26):\penalty0
  2921--2924, December 1988.
\newblock \doi{10.1103/PhysRevLett.61.2921}.

\bibitem[Greenberger et~al.(1989)Greenberger, Horne, and
  Zeilinger]{Greenberger.Horne.ea:1989}
Daniel~M. Greenberger, Michael~A. Horne, and Anton Zeilinger.
\newblock Going {{Beyond Bell}}'s {{Theorem}}.
\newblock In Menas Kafatos, editor, \emph{Bell's {{Theorem}}, {{Quantum
  Theory}} and {{Conceptions}} of the {{Universe}}}, Fundamental {{Theories}}
  of {{Physics}}, pages 69--72. {Springer Netherlands}, {Dordrecht}, 1989.
\newblock ISBN 978-94-017-0849-4.
\newblock \doi{10.1007/978-94-017-0849-4_10}.

\bibitem[Briegel and Raussendorf(2001)]{Briegel.Raussendorf:2001}
Hans~J. Briegel and Robert Raussendorf.
\newblock Persistent {{Entanglement}} in {{Arrays}} of {{Interacting
  Particles}}.
\newblock \emph{Physical Review Letters}, 86\penalty0 (5):\penalty0 910--913,
  January 2001.
\newblock \doi{10.1103/PhysRevLett.86.910}.

\bibitem[Brown and Twiss(1956)]{Brown.Twiss:1956}
R.~Hanbury Brown and R.~Q. Twiss.
\newblock Correlation between {{Photons}} in two {{Coherent Beams}} of
  {{Light}}.
\newblock \emph{Nature}, 177\penalty0 (4497):\penalty0 27--29, January 1956.
\newblock \doi{10.1038/177027a0}.

\bibitem[Steudle et~al.(2012)Steudle, Schietinger, H{\"o}ckel, Dorenbos, Zadeh,
  Zwiller, and Benson]{Steudle.Schietinger.ea:2012}
Gesine~A. Steudle, Stefan Schietinger, David H{\"o}ckel, Sander~N. Dorenbos,
  Iman~E. Zadeh, Valery Zwiller, and Oliver Benson.
\newblock Measuring the quantum nature of light with a single source and a
  single detector.
\newblock \emph{Physical Review A}, 86\penalty0 (5):\penalty0 053814, November
  2012.
\newblock \doi{10.1103/PhysRevA.86.053814}.

\bibitem[Roberts and {Ali-Bakhshian}(2010)]{Roberts.Ali-Bakhshian:2010}
Gordon~W. Roberts and Mohammad {Ali-Bakhshian}.
\newblock A {{Brief Introduction}} to {{Time}}-to-{{Digital}} and
  {{Digital}}-to-{{Time Converters}}.
\newblock \emph{IEEE Transactions on Circuits and Systems II: Express Briefs},
  57\penalty0 (3):\penalty0 153--157, March 2010.
\newblock \doi{10.1109/TCSII.2010.2043382}.

\bibitem[Zopf et~al.(2019)Zopf, Keil, Chen, Yang, Chen, Ding, and
  Schmidt]{Zopf.Keil.ea:2019}
Michael Zopf, Robert Keil, Yan Chen, Jingzhong Yang, Disheng Chen, Fei Ding,
  and Oliver~G. Schmidt.
\newblock Entanglement {{Swapping}} with {{Semiconductor}}-{{Generated Photons
  Violates Bell}}'s {{Inequality}}.
\newblock \emph{Physical Review Letters}, 123\penalty0 (16):\penalty0 160502,
  October 2019.
\newblock \doi{10.1103/PhysRevLett.123.160502}.

\bibitem[Basso~Basset et~al.(2019)Basso~Basset, Rota, Schimpf, Tedeschi,
  Zeuner, {Covre da Silva}, Reindl, Zwiller, J{\"o}ns, Rastelli, and
  Trotta]{BassoBasset.Rota.ea:2019}
F.~Basso~Basset, M.~B. Rota, C.~Schimpf, D.~Tedeschi, K.~D. Zeuner, S.~F.
  {Covre da Silva}, M.~Reindl, V.~Zwiller, K.~D. J{\"o}ns, A.~Rastelli, and
  R.~Trotta.
\newblock Entanglement {{Swapping}} with {{Photons Generated}} on {{Demand}} by
  a {{Quantum Dot}}.
\newblock \emph{Physical Review Letters}, 123\penalty0 (16):\penalty0 160501,
  October 2019.
\newblock \doi{10.1103/PhysRevLett.123.160501}.

\bibitem[Ursin et~al.(2007)Ursin, Tiefenbacher, {Schmitt-Manderbach}, Weier,
  Scheidl, Lindenthal, Blauensteiner, Jennewein, Perdigues, Trojek, {\"O}mer,
  F{\"u}rst, Meyenburg, Rarity, Sodnik, Barbieri, Weinfurter, and
  Zeilinger]{Ursin.Tiefenbacher.ea:2007}
R.~Ursin, F.~Tiefenbacher, T.~{Schmitt-Manderbach}, H.~Weier, T.~Scheidl,
  M.~Lindenthal, B.~Blauensteiner, T.~Jennewein, J.~Perdigues, P.~Trojek,
  B.~{\"O}mer, M.~F{\"u}rst, M.~Meyenburg, J.~Rarity, Z.~Sodnik, C.~Barbieri,
  H.~Weinfurter, and A.~Zeilinger.
\newblock Entanglement-based quantum communication over 144 km.
\newblock \emph{Nature Physics}, 3\penalty0 (7):\penalty0 481--486, July 2007.
\newblock \doi{10.1038/nphys629}.

\bibitem[Shalm et~al.(2013)Shalm, Hamel, Yan, Simon, Resch, and
  Jennewein]{Shalm.Hamel.ea:2013}
L.~K. Shalm, D.~R. Hamel, Z.~Yan, C.~Simon, K.~J. Resch, and T.~Jennewein.
\newblock Three-photon energy\textendash time entanglement.
\newblock \emph{Nature Physics}, 9\penalty0 (1):\penalty0 19--22, January 2013.
\newblock \doi{10.1038/nphys2492}.

\bibitem[Delteil et~al.(2016)Delteil, Sun, Gao, Togan, Faelt, and Imamo{\u
  g}lu]{Delteil.Sun.ea:2016}
Aymeric Delteil, Zhe Sun, Wei-bo Gao, Emre Togan, Stefan Faelt, and Ata{\c c}
  Imamo{\u g}lu.
\newblock Generation of heralded entanglement between distant hole spins.
\newblock \emph{Nature Physics}, 12\penalty0 (3):\penalty0 218--223, March
  2016.
\newblock \doi{10.1038/nphys3605}.

\bibitem[Reindl et~al.(2018)Reindl, Huber, Schimpf, da~Silva, Rota, Huang,
  Zwiller, J{\"o}ns, Rastelli, and Trotta]{Reindl.Huber.ea:2018}
Marcus Reindl, Daniel Huber, Christian Schimpf, Saimon F.~Covre da~Silva,
  Michele~B. Rota, Huiying Huang, Val Zwiller, Klaus~D. J{\"o}ns, Armando
  Rastelli, and Rinaldo Trotta.
\newblock All-photonic quantum teleportation using on-demand solid-state
  quantum emitters.
\newblock \emph{Science Advances}, 4\penalty0 (12):\penalty0 eaau1255, December
  2018.
\newblock \doi{10.1126/sciadv.aau1255}.

\bibitem[Sch{\"o}ll et~al.(2019)Sch{\"o}ll, Hanschke, Schweickert, Zeuner,
  Reindl, {Covre da Silva}, Lettner, Trotta, Finley, M{\"u}ller, Rastelli,
  Zwiller, and J{\"o}ns]{Scholl.Hanschke.ea:2019}
Eva Sch{\"o}ll, Lukas Hanschke, Lucas Schweickert, Katharina~D. Zeuner, Marcus
  Reindl, Saimon~Filipe {Covre da Silva}, Thomas Lettner, Rinaldo Trotta,
  Jonathan~J. Finley, Kai M{\"u}ller, Armando Rastelli, Val Zwiller, and
  Klaus~D. J{\"o}ns.
\newblock Resonance {{Fluorescence}} of {{GaAs Quantum Dots}} with
  {{Near}}-{{Unity Photon Indistinguishability}}.
\newblock \emph{Nano Letters}, 19\penalty0 (4):\penalty0 2404--2410, April
  2019.
\newblock \doi{10.1021/acs.nanolett.8b05132}.

\bibitem[GmbH(2020)]{PicoQuantGmbH:2020}
PicoQuant GmbH.
\newblock Software - {{Time}}-{{Resolved Fluoresence Wiki}}.
\newblock https://perma.cc/FGT9-RD7T, September 2020.

\bibitem[Aharonovich et~al.(2016)Aharonovich, Englund, and
  Toth]{Aharonovich.Englund.ea:2016}
Igor Aharonovich, Dirk Englund, and Milos Toth.
\newblock Solid-state single-photon emitters.
\newblock \emph{Nature Photonics}, 10\penalty0 (10):\penalty0 631--641, October
  2016.
\newblock \doi{10.1038/nphoton.2016.186}.

\bibitem[Becker(2012)]{Becker:2012}
W.~Becker.
\newblock Fluorescence lifetime imaging \textendash{} techniques and
  applications.
\newblock \emph{Journal of Microscopy}, 247\penalty0 (2):\penalty0 119--136,
  2012.
\newblock \doi{10.1111/j.1365-2818.2012.03618.x}.

\bibitem[Buller and Wallace(2007)]{Buller.Wallace:2007}
Gerald Buller and Andrew Wallace.
\newblock Ranging and {{Three}}-{{Dimensional Imaging Using Time}}-{{Correlated
  Single}}-{{Photon Counting}} and {{Point}}-by-{{Point Acquisition}}.
\newblock \emph{IEEE Journal of Selected Topics in Quantum Electronics},
  13\penalty0 (4):\penalty0 1006--1015, July 2007.
\newblock \doi{10.1109/JSTQE.2007.902850}.

\bibitem[Freedman and Clauser(1972)]{Freedman.Clauser:1972}
Stuart~J. Freedman and John~F. Clauser.
\newblock Experimental {{Test}} of {{Local Hidden}}-{{Variable Theories}}.
\newblock \emph{Physical Review Letters}, 28\penalty0 (14):\penalty0 938--941,
  April 1972.
\newblock \doi{10.1103/PhysRevLett.28.938}.

\bibitem[De~Greve et~al.(2012)De~Greve, Yu, McMahon, Pelc, Natarajan, Kim, Abe,
  Maier, Schneider, Kamp, H{\"o}fling, Hadfield, Forchel, Fejer, and
  Yamamoto]{DeGreve.Yu.ea:2012}
Kristiaan De~Greve, Leo Yu, Peter~L. McMahon, Jason~S. Pelc, Chandra~M.
  Natarajan, Na~Young Kim, Eisuke Abe, Sebastian Maier, Christian Schneider,
  Martin Kamp, Sven H{\"o}fling, Robert~H. Hadfield, Alfred Forchel, M.~M.
  Fejer, and Yoshihisa Yamamoto.
\newblock Quantum-dot spin\textendash photon entanglement via frequency
  downconversion to telecom wavelength.
\newblock \emph{Nature}, 491\penalty0 (7424):\penalty0 421--425, November 2012.
\newblock \doi{10.1038/nature11577}.

\bibitem[Gao et~al.(2012)Gao, Fallahi, Togan, {Miguel-Sanchez}, and
  Imamoglu]{Gao.Fallahi.ea:2012}
W.~B. Gao, P.~Fallahi, E.~Togan, J.~{Miguel-Sanchez}, and A.~Imamoglu.
\newblock Observation of entanglement between a quantum dot spin and a single
  photon.
\newblock \emph{Nature}, 491\penalty0 (7424):\penalty0 426--430, November 2012.
\newblock \doi{10.1038/nature11573}.

\bibitem[B{\"o}hmer et~al.(2001)B{\"o}hmer, Pampaloni, Wahl, Rahn, Erdmann, and
  Enderlein]{Bohmer.Pampaloni.ea:2001}
Martin B{\"o}hmer, Francesco Pampaloni, Michael Wahl, Hans-J{\"u}rgen Rahn,
  Rainer Erdmann, and J{\"o}rg Enderlein.
\newblock Time-resolved confocal scanning device for ultrasensitive
  fluorescence detection.
\newblock \emph{Review of Scientific Instruments}, 72\penalty0 (11):\penalty0
  4145--4152, November 2001.
\newblock \doi{10.1063/1.1406926}.

\bibitem[Bennett and Brassard(1984)]{Bennett.Brassard:1984}
Charles Bennett and Gilles Brassard.
\newblock Quantum cryptography: {{Public}} key distribution and coin tossing.
\newblock In \emph{Proceedings of {{IEEE International Conference}} on
  {{Computers}}, {{Systems}} and {{Signal Processing}}}, volume 175, page~8,
  {Bangelore, India}, January 1984.
\newblock \doi{10.1016/j.tcs.2011.08.039}.

\bibitem[Yin et~al.(2020)Yin, Li, Liao, Yang, Cao, Zhang, Ren, Cai, Liu, Li,
  Shu, Huang, Deng, Li, Zhang, Liu, Chen, Lu, Wang, Xu, Wang, Peng, Ekert, and
  Pan]{Yin.Li.ea:2020}
Juan Yin, Yu-Huai Li, Sheng-Kai Liao, Meng Yang, Yuan Cao, Liang Zhang, Ji-Gang
  Ren, Wen-Qi Cai, Wei-Yue Liu, Shuang-Lin Li, Rong Shu, Yong-Mei Huang, Lei
  Deng, Li~Li, Qiang Zhang, Nai-Le Liu, Yu-Ao Chen, Chao-Yang Lu, Xiang-Bin
  Wang, Feihu Xu, Jian-Yu Wang, Cheng-Zhi Peng, Artur~K. Ekert, and Jian-Wei
  Pan.
\newblock Entanglement-based secure quantum cryptography over 1,120 kilometres.
\newblock \emph{Nature}, 582\penalty0 (7813):\penalty0 501--505, June 2020.
\newblock \doi{10.1038/s41586-020-2401-y}.

\bibitem[Knuth(1968)]{Knuth:1968}
Donald Knuth.
\newblock \emph{The {{Art}} of {{Computer Programming}}}.
\newblock {Addison-Wesley}, 1968.
\newblock ISBN 0-201-03801-3.

\bibitem[Hoare(1961)]{Hoare:1961}
C.~A.~R. Hoare.
\newblock Algorithm 64: {{Quicksort}}.
\newblock \emph{Communications of the ACM}, 4\penalty0 (7):\penalty0 321, July
  1961.
\newblock \doi{10.1145/366622.366644}.

\bibitem[Bischof(2012)]{Bischof:2012}
Thomas Bischof.
\newblock Photon {{Correlation}}, February 2012.
\newblock URL \url{https://github.com/tsbischof/photon_correlation}.

\bibitem[Ballesteros et~al.(2019)Ballesteros, Proux, Bonato, and
  Gerardot]{Ballesteros.Proux.ea:2019}
G.~C. Ballesteros, R.~Proux, C.~Bonato, and B.~D. Gerardot.
\newblock {{readPTU}}: A python library to analyse time tagged time resolved
  data.
\newblock \emph{Journal of Instrumentation}, 14\penalty0 (06):\penalty0
  T06011--T06011, June 2019.
\newblock \doi{10.1088/1748-0221/14/06/T06011}.

\bibitem[Lam et~al.(2015)Lam, Pitrou, and Seibert]{Lam.Pitrou.ea:2015}
Siu~Kwan Lam, Antoine Pitrou, and Stanley Seibert.
\newblock Numba: A {{LLVM}}-based {{Python JIT}} compiler.
\newblock In \emph{Proceedings of the {{Second Workshop}} on the {{LLVM
  Compiler Infrastructure}} in {{HPC}} - {{LLVM}} '15}, pages 1--6, {Austin,
  Texas}, 2015. {ACM Press}.
\newblock ISBN 978-1-4503-4005-2.
\newblock \doi{10.1145/2833157.2833162}.

\bibitem[Lattner and Adve(2004)]{Lattner.Adve:2004}
Chris Lattner and Vikram Adve.
\newblock {{LLVM}}: {{A Compilation Framework}} for {{Lifelong Program
  Analysis}} \& {{Transformation}}.
\newblock In \emph{Proceedings of the International Symposium on {{Code}}
  Generation and Optimization: Feedback-Directed and Runtime Optimization},
  {{CGO}} '04, page~75, {USA}, March 2004. {IEEE Computer Society}.
\newblock ISBN 978-0-7695-2102-2.

\bibitem[Kimble et~al.(1977)Kimble, Dagenais, and
  Mandel]{Kimble.Dagenais.ea:1977}
H.~J. Kimble, M.~Dagenais, and L.~Mandel.
\newblock Photon {{Antibunching}} in {{Resonance Fluorescence}}.
\newblock \emph{Physical Review Letters}, 39\penalty0 (11):\penalty0 691--695,
  September 1977.
\newblock \doi{10.1103/PhysRevLett.39.691}.

\bibitem[Magde et~al.(1972)Magde, Elson, and Webb]{Magde.Elson.ea:1972}
Douglas Magde, Elliot Elson, and W.~W. Webb.
\newblock Thermodynamic {{Fluctuations}} in a {{Reacting
  System}}---{{Measurement}} by {{Fluorescence Correlation Spectroscopy}}.
\newblock \emph{Physical Review Letters}, 29\penalty0 (11):\penalty0 705--708,
  September 1972.
\newblock \doi{10.1103/PhysRevLett.29.705}.

\bibitem[Moerner and Fromm(2003)]{Moerner.Fromm:2003}
W.~E. Moerner and David~P. Fromm.
\newblock Methods of single-molecule fluorescence spectroscopy and microscopy.
\newblock \emph{Review of Scientific Instruments}, 74\penalty0 (8):\penalty0
  3597--3619, July 2003.
\newblock \doi{10.1063/1.1589587}.

\bibitem[Dickson et~al.(1997)Dickson, Cubitt, Tsien, and
  Moerner]{Dickson.Cubitt.ea:1997}
Robert~M. Dickson, Andrew~B. Cubitt, Roger~Y. Tsien, and W.~E. Moerner.
\newblock On/off blinking and switching behaviour of single molecules of green
  fluorescent protein.
\newblock \emph{Nature}, 388\penalty0 (6640):\penalty0 355--358, July 1997.
\newblock \doi{10.1038/41048}.

\bibitem[Davidson and Mandel(1968)]{Davidson.Mandel:1968}
F.~Davidson and L.~Mandel.
\newblock Photoelectric {{Correlation Measurements}} with
  {{Time}}-to-{{Amplitude Converters}}.
\newblock \emph{Journal of Applied Physics}, 39\penalty0 (1):\penalty0 62--66,
  January 1968.
\newblock \doi{10.1063/1.1655781}.

\bibitem[Dean and Ghemawat(2008)]{Dean.Ghemawat:2008}
Jeffrey Dean and Sanjay Ghemawat.
\newblock {{MapReduce}}: Simplified data processing on large clusters.
\newblock \emph{Communications of the ACM}, 51\penalty0 (1):\penalty0 107--113,
  January 2008.
\newblock \doi{10.1145/1327452.1327492}.

\bibitem[Wang et~al.(2019)Wang, Qin, Ding, Chen, Chen, You, He, Jiang, You,
  Wang, Schneider, Renema, Höfling, Lu, and Pan]{Wang.Qin.ea:2019}
Hui Wang, Jian Qin, Xing Ding, Ming-Cheng Chen, Si~Chen, Xiang You, Yu-Ming He,
  Xiao Jiang, L.~You, Z.~Wang, C.~Schneider, Jelmer~J. Renema, Sven Höfling,
  Chao-Yang Lu, and Jian-Wei Pan.
\newblock Boson sampling with 20 input photons and a 60-mode interferometer in
  a {$1{0}^{14}$--Dimensional} hilbert space.
\newblock \emph{Physical Review Letters}, 123\penalty0 (25):\penalty0 250503,
  December 2019.
\newblock \doi{10.1103/PhysRevLett.123.250503}.

\bibitem[Knill et~al.(2001)Knill, Laflamme, and
  Milburn]{Knill.Laflamme.ea:2001}
E.~Knill, R.~Laflamme, and G.~J. Milburn.
\newblock A scheme for efficient quantum computation with linear optics.
\newblock \emph{Nature}, 409\penalty0 (6816):\penalty0 46--52, January 2001.
\newblock \doi{10.1038/35051009}.

\end{thebibliography}
\end{document}